\documentclass[aps,prl,twocolumn]{revtex4-2}
\usepackage{mathtools}
\usepackage{xcolor}
\usepackage{graphicx}
\usepackage{amsmath}
\usepackage{amssymb}

\newcommand{\av}[1]{\left\langle #1 \right\rangle}

\begin{document}

\title{Dynamics of oscillator populations globally coupled with distributed phase shifts}

\author{Lev A. Smirnov}
\affiliation{Department of Control Theory, Research and Education Mathematical Center ``Mathematics for Future Technologies'',
Nizhny Novgorod State University, Gagarin Av. 23, 603022, Nizhny Novgorod, Russia}
\author{Arkady Pikovsky}
\affiliation{Department of Physics and Astronomy, University of Potsdam, Karl-Liebknecht-Str. 24/25, 14476, Potsdam-Golm, Germany}


\begin{abstract}
We consider a population of globally coupled oscillators in which phase shifts in the coupling are random. 
We show that in the maximally disordered case, where the pairwise shifts are i.i.d. random variables, the dynamics of a large population reduces to one without randomness in the shifts but with an effective coupling function, which is a convolution of the original coupling function with the distribution of the phase shifts.
This result is valid for noisy oscillators and/or in presence of a distribution of natural frequencies.
We argue also, using the property of global asymptotic stability, that this reduction is valid in  a partially disordered case, where random phase shifts are attributed to the forced units only.
However, the reduction to an effective coupling in the partially disordered noise-free situation may fail if the coupling function is complex enough to ensure the multistability of locked states.
\end{abstract}

\maketitle

Globally coupled populations of oscillators serve, since works of Winfree and Kuramoto~\cite{Winfree-67,Kuramoto-75}, as paradigmatic models for collective synchronization.
They describe, in particular, arrays of Josephson junctions~\cite{Wiesenfeld-Colet-Strogatz-98}, coupled spin-torque, micromechanical and electrochemical oscillators~\cite{Kiss-Zhai-Hudson-02a,Tiberkevich_etal-09,Heinrich_etal-11}, Belousov-Zhabotinsky chemical oscillators in droplets~\cite{toiya2010synchronization}.
In many cases, the global nature of coupling is determined by the setup: a common load for Josephson junction, spin-torque  or electrochemical oscillators naturally ensures that a common oscillatory current flows through these units.
Also, mechanical oscillators (metronomes or pedestrians) on a common platform experience the same field.
However, the oscillators may have different properties and possess intrinsic noise so that despite a global force, the equations for them are not identical.

The most common source of diversity is a spread of natural frequencies of oscillators; this feature has been incorporated already in the original Kuramoto model~\cite{Kuramoto-75}.
Later, one generalized this for a disorder in other parameters of the oscillators~\cite{Pazo-Montbrio-11,Iatsenko_etal-13,Vlasov-Macau-Pikovsky-14,lee2009large}.
In this letter, we focus on the effect of a disorder in the phase shifts in the coupling.
The phase shift was absent in the original Kuramoto setup~\cite{Kuramoto-75}, but was introduced in a subsequent paper by Sakaguchi and Kuramoto~\cite{Sakaguchi-Kuramoto-86}.
A constant (the same for all units) phase shift in the coupling, taken into account by Sakaguchi and Kuramoto, naturally appears when collective modes are responsible for the interaction (e.g., macroscopic oscillations of a common load for the Josephson junction~\cite{Wiesenfeld-Colet-Strogatz-98} or of a platform  for the metronomes~\cite{kapitaniak2012synchronization}) are phase-shifted with respect to the forcing of the oscillators. 

We study here a situation where the phase shifts $\alpha$ for the oscillators in the population are different.
We start with a situation of maximal disorder, where all the phase shifts $\alpha_{jk}$ between interacting units $j,k$ are independent random variables~\cite{martens2016chimera} obeying a distribution with density
\begin{equation}
g(\alpha)=\frac{1}{2\pi}\sum_m \eta_me^{im\alpha}, \quad
\eta_m=\int_0^{2\pi}\!\!d\alpha\,e^{-im\alpha}g(\alpha)\,.
\label{eq:ga}
\end{equation}
We consider a rather generic setup which includes noise and possible heterogeneity of oscillators 
\begin{equation}
\dot\varphi_k=F_k(\varphi_{k})+\sigma\xi_k(t)+\varepsilon\frac{1}{N}\sum_j \Gamma(\varphi_j-\varphi_k-\alpha_{jk})\;.
\label{eq:dk}
\end{equation}
Here the term $F_k$ describes local dynamics of uncoupled phases; for oscillators it is usually just the natural frequency $F_k\!=\!\omega_k$, but it can take more complex form for active rotators~\cite{Sakaguchi-Kuramoto-86,klinshov2021effect} or theta-neurons~\cite{luke2013complete}.
Furthermore, we include independent white Gaussian noise terms $~\sigma \xi_k(t)$, $\av{\xi_k(t)\xi_m(t')=2\delta_{km}\delta(t-t')}$.
The generic Daido-Kuramoto coupling term~\cite{Daido-96} $\varepsilon\Gamma$ is defined by a $2\pi$-periodic coupling function $\Gamma(x)=\sum_m f_me^{im x}$.
We keep a small parameter $\varepsilon$ to stress that this model is valid for weakly coupled oscillators in the first order in this parameter. 

The nature in the disorder in phase shifts $\alpha_{jk}$ can be manifold, because to these phase shifts contribute the forcing unit, the transmission from the driving to the driven unit, and the phase sensitivity property of the forced unit. 
For example, a transmission of a signal from the driving unit to the driven one can possess time delays~\cite{lee2009large}, resulting in a spread of the phase shifts (such delays are crucial for the brain dynamics~\cite{deco2009key}).
The equivalence of time delays to phase shifts for weakly coupled oscillators has been established in~\cite{Izhikevich-98}: $\alpha_{jk}\!=\!\omega_0\tau_{jk}$, where $\omega_0$ is the characteristic frequency of the oscillators and $\tau_{jk}$ are the delays.
This equivalence heavily relies on the existence of the small parameter $\varepsilon$.
First, although the characteristic frequencies of the oscillators and of the mean field may differ from $\omega_0$ in order $\varepsilon$, these effects can be neglected if we keep only the order $\sim\varepsilon$ in the coupling term.
The second condition is that the delays should be smaller than the slow time scale, i.e. $\varepsilon\tau_{jk}\ll{1}$.
Because parameter $\varepsilon$ is small, the phase shifts $\omega_0\tau_{jk}$ can be still of order $2\pi$ or larger. 

Fourier representation of the coupling function $\Gamma$ allows for rewriting the coupling term as
\begin{equation}
\!\frac{1}{N}\!\sum_j\Gamma(\varphi_j-\varphi_k-\alpha_{jk})\!=\!
\sum_m\!f_m e^{-im\varphi_k}\frac{1}{N}\!\sum_j\!e^{im\varphi_j-im\alpha_{jk}}.
\label{eq:sum}
\end{equation}
We identify the last sum over index $j$ as an average over the ensemble, and to calculate it, we make a crucial assumption on the statistical independence of the phases and the phase shifts.
This assumption appears reasonable in the thermodynamic limit $N\!\to\!\infty$, where an oscillator is subject to a sum of many random forces.
A similar assumption about independence of the order parameters and random time delays has been made in~\cite{lee2009large}. Thus,
\begin{equation}
\frac{1}{N}\sum_j
e^{im\varphi_j-im\alpha_{jk}}=\langle e^{im\varphi}\rangle \langle e^{-im\alpha}\rangle=Z_m\eta_m\;,
\label{eq:dec}
\end{equation}
where we introduce the standard Kuarmoto-Daido order parameters $Z_m\!=\!\langle e^{im\varphi}\rangle$.
Substituting back in~\eqref{eq:sum} we obtain a new effective coupling which does not contain random phase shifts
\begin{gather}
\frac{1}{N}\sum_j\sum_m f_m\eta_me^{im(\varphi_j-\varphi_k)}=\frac{1}{N}\sum_j\tilde\Gamma(\varphi_j-\varphi_k)\;,
\label{eq:nc}\\
\tilde\Gamma(x)=\int_0^{2\pi}\!\!d\alpha\,\Gamma(x-\alpha)g(\alpha)\;.
\label{eq:conv}
\end{gather}
The effective coupling is just the convolution of the original one with the distribution density of the phase shifts.
This effective coupling then can be used in Eq.~\eqref{eq:dk}.
This reduction for the case of maximal disorder is the main result of the first part of this letter.
Expression~\eqref{eq:conv} and the equivalent expression for the Fourier modes $\tilde f_m=f_m\eta_m$ shows that the coupling becomes weaker and, more important, its form changes.
The ``simplification'' of the coupling is mostly pronounced if the distribution of the phase shifts has only one harmonics, e.g., the first harmonics $\eta_1\neq 0$, $\eta_{n>1}=0$:
\begin{equation}
g(\alpha)=(2\pi)^{-1}\Bigl(1+2|\eta_1| \cos(\alpha-\arg(\eta_1)\Bigr)\;.
\label{eq:g1h}
\end{equation}
(The case of a Gamma-distribution of the time delays considered in~\cite{lee2009large} is less instructive, because there $\eta_m\neq 0$ for all $m$.)
For such a distribution of phase shifts, the effective coupling is just a $\sin$-coupling, independently of the form of the original function $\Gamma$ (provided it contains a nonvanishing first harmonics).
Remarkably, for a pure $\sin$-coupling, many analytical results are available: for noiseless oscillators with a Cauchy distribution of natural frequencies $\omega_k$, the Ott-Antonsen (OA) ansatz~\cite{Ott-Antonsen-08} is valid; for oscillators with identical natural frequencies and with Gaussian noise, there is an analytical solution for $Z_1$ in dependence on the coupling strength and noise in terms of modified Bessel functions~\cite{Munyaev_etal_2020}.
We have checked in numerical simulations, that indeed these relations become valid if one considers ensembles with complex coupling functions $\Gamma$ but with a maximally simplifying distribution of phase shifts~\eqref{eq:g1h}.

The reduction above is quite generic, but it relies on the assumption of independence of the phases and the phase shifts, for which we have only plausible arguments.
Below we consider the case of partial disorder, where in some situations the arguments supporting the reduction \eqref{eq:nc},\eqref{eq:conv} can be made rather accurate, but the reduction will be valid not in all cases.
By partial disorder we denote situations where the phase shifts in the coupling depend on one index only: $\alpha_{jk}=\alpha_j$, or $\alpha_{jk}=\alpha_k$.
The interpretation of these cases is straightforward: in the former situation the phase shift is the property of the driving unit (see examples in~\cite{maistrenko2014solitary}), in the latter case the phase shift is attributed to the receiver,
like in Refs.~\cite{Hong-Strogatz-11,Vlasov-Macau-Pikovsky-14,iatsenko2014glassy}.
For example, in~\cite{Hong-Strogatz-11} the authors consider two types of oscillators: conformists and contrarians.
In terms of the phase shifts, conformists adjust to the driving field and have $\alpha_k\!=\!0$, while contrarians prefer to be out of phase and have $\alpha_k\!=\!\pi$.
Another situation considered in~\cite{Vlasov-Macau-Pikovsky-14,Vlasov-Pikovsky-Macau-15} is when a global driving field is ``broadcast'' from one source, and a unit $\varphi_k$ receives this forcing with delay $\tau_k$ proportional to the distance to the ``loudspeaker''.
According to the equivalence discussed above, the distribution of positions of the units results in a distribution of the phase shifts according to $\alpha_k=\omega_0\tau_k$.
Remarkably, the two cases (dependence on index $j$ or on index $k$) can be transformed to each other by a simple shift of the variables~\cite{Vlasov-Macau-Pikovsky-14,iatsenko2014glassy}.
Thus we consider the latter case only.

Now the statistical independence assumption can be hardly justified, therefore we apply other methods, for two particular situations: oscillators with an arbitrary distribution of natural frequencies $\omega_k$ and external noise $\sim\!\sigma$; and noise-free oscillators with $\sin$-coupling and Lorentzian distribution of frequencies, where we can apply the OA theory.
We write the equations of the first model using \eqref{eq:dk} with $F_k=\omega_k$ and the mode representation of $\Gamma$
\begin{equation}
\dot\varphi_k=\omega_k+\sum_m f_m Z_me^{-im(\varphi_k+\alpha_k)}+\sigma\xi_k(t)\;.
\label{eq:gkm}
\end{equation}
In the thermodynamic limit, we describe the evolution of the probability density $P(\varphi,t|\omega,\alpha)$ of the phases $\varphi$, conditioned by the phase shifts $\alpha$ and frequencies $\omega$, with the Fokker-Planck equation
\begin{equation}
\partial_t P+\partial_\varphi \biggl[\biggl(\omega+\sum_m f_m Z_m e^{-im(\varphi+\alpha)}\biggr)P\biggr]\!=\sigma^2\partial_{\varphi\varphi}P\;.
\label{eq:fpe}
\end{equation}
The complex order parameters are represented as $Z_m(t)\!=\!\int d\omega \,w(\omega)\int_0^{2\pi}\!d\varphi\int_0^{2\pi}\!d\alpha\, g(\alpha) e^{im\varphi} P(\varphi,t|\omega,\alpha)$, where{\,}$w(\omega)${\,}is a probability density function of frequencies.
We now change the variable $\theta\!=\!\varphi\!+\!\alpha$ and for $P(\theta,t|\omega,\alpha)${\,}obtain
\begin{equation}
\partial_t P+\partial_\theta \bigg[\biggl(\omega+\sum_m f_m Z_m e^{-im \theta}\biggr)P\biggr]=\sigma^2\partial_{\theta\theta}P\;.
\label{eq:th}
\end{equation}
This change of the variable effectively ``moves'' the phase shifts from the dynamical equations to the definition of the complex order parameters: $Z_m\!=\!\int d\omega\, w(\omega)\int_0^{2\pi}\! d\theta \int_0^{2\pi} \!d\alpha\, g(\alpha)\, e^{im\theta-im\alpha} P(\theta,t|\omega,\alpha)$.
The crucial observation is that although the density $P(\theta,t|\omega,\alpha)$ generally depends on the phase shifts, Eq.~\eqref{eq:th} does not.
We now argue that any initial  dependence of $P$ on the parameter $\alpha$ will eventually disappear, so that asymptotically at large times  $P(\theta,t|\omega,\alpha)\to P(\theta,t|\omega)$.
This property is well-established for a time-independent Fokker-Planck equation~\cite{Gardiner-96}.
In terms of the theory of partial differential equations, this property means global asymptotic stability (GAS) of solutions of the parabolic Fokker-Planck equation~\eqref{eq:th}, cf.~\cite{Calogero-12}.
In the context of general Markov processes, this property is nothing else as ergodicity, often formulated as ``loss of memory''~\cite{Kulik-18}.
Although GAS is expected to be valid for~\eqref{eq:th}~\cite{KK}, we have not found a proof in the mathematical literature.
A proof of GAS for the time-dependent master equation, which is a closest finite-dimensional analogue of the Fokker-Planck equation, has been given in Ref.~\cite{Earnshaw-Keener-10}.
From the physical viewpoint, GAS of Eq.~\eqref{eq:th} appears rather evident, particularly because the equation is defined on a finite domain $0\leq\theta<2\pi$, and one does not face difficulties in defining a convergence of probability distributions on an infinite domain.

With the GAS property, for large times, the order parameters $Z_m$ can be expressed via the density{\,}$P(\theta,t|\omega)${\,}as
\begin{equation}
\begin{gathered}
Z_m(t)=\eta_m\int\!d\omega\,w(\omega) \int_0^{2\pi}\!\!d\theta \,e^{im\theta} P(\theta,t|\omega)\;.
\end{gathered}
\label{eq:orp}
\end{equation}
As a result of substituting this in~\eqref{eq:th}, we obtain the same reduction~\eqref{eq:nc},{\,}\eqref{eq:conv} as for the maximal disorder above: one reduces the system to one without phase shifts, but  the order parameters of this auxiliary population have to be multiplied, according to~\eqref{eq:orp}, with the circular moments $\eta_m$ of the distribution of the phase shifts $g(\alpha)$. 
Or, equivalently, one can say that the phases $\theta$ obey the dynamics of the globally coupled ensemble with the effective coupling function~\eqref{eq:conv}.

\looseness=-1
We stress here that while the full problem~\eqref{eq:fpe} as a \textit{nonlinear} Fokker-Planck equation can demonstrate multistability and hysteresis, this does not contradict GAS because, for the latter property, one considers Eq.~\eqref{eq:th} with predetermined values of the moments $Z_m(t)$ (in other words, GAS corresponds to a transversal stability of a solution for a particular time course of the global force).

The property of a unique asymptotic solution of the auxiliary problem heavily
relies on the loss of memory (dissipativity due to noise-induced diffusion) of the kinetic equation \eqref{eq:th}.
We next demonstrate that this dissipativity also may occur in a noise-free setup due to a distribution of natural frequencies in the population.
For this, we have to assume additionally that a distribution of natural frequencies $w(\omega)$ is independent of the distribution of phase shifts $g(\alpha)$.
Furthermore, to be able to apply the OA approach~\cite{Ott-Antonsen-08}, we assume that the coupling is due to the first harmonics only, i.e. $f_m\!=\!0$ for $|m|\!>\!1$.
Under these assumptions, the kinetic equation for the density $P(\theta,t|\omega,\alpha)$ (which is the same as~\eqref{eq:th} but with $\sigma\!=\!0$) can be represented through an infinite set of equations for the $\omega,\alpha$-dependent order parameters $q_m(t|\omega,\alpha)=\int_0^{2\pi}\!d\theta\,e^{im\theta}P(\theta,t|\omega,\alpha)$.
The main order parameters are integrals of these quantities $Z_m(t)\!=\!\int\!d\alpha\, g(\alpha) \int\!d\omega \,w(\omega)\,q_m(t|\omega,\alpha)$.
In the OA theory, one first assumes analyticity in the upper half plane of these order parameters as functions of the frequency.
This allows, provided the distribution $w(\omega)$ is a Cauchy one $w_c(\omega)\!=\!\gamma\bigl[\pi\bigl((\omega-\omega_0)^2+\gamma^2\bigr)\bigr]^{-1}$, for integration over frequencies by virtue of the residue theorem $\int\! d\omega \,w_c(\omega)\,q_m(t|\omega,\alpha)\!=\!q_m(t|\omega_0+i\gamma,\alpha)$.
The second constituent of the OA theory is the observation, that the infinite system of equations for $q_m$ possesses an invariant manifold $q_m=Q^m$, corresponding to a wrapped Cauchy distribution of the phases $\theta$.
Thus, the whole hierarchy of equations reduces just to one equation (where $\alpha$ is still a parameter, and we denote $H=f_1Z_1$)
\begin{gather}
\dot Q(t|\alpha)=(i\omega_0-\gamma) Q(t|\alpha)+H-H^* Q^2 (t|\alpha)\;, \label{eq:eqQ}\\
Z_m(t)=\int_0^{2\pi}\!\! d\alpha\, Q^m(t|\alpha) e^{-im\alpha} g(\alpha)\;.\label{eq:eqZ}
\end{gather}

Note that in Eq.~\eqref{eq:eqQ} the order parameter $Q(t|\alpha)$ depends on the phase shifts $\alpha$ only through initial conditions, but its dynamics is $\alpha$-independent.
The GAS property above means that in the course of the dynamics, the memory about the initial state gets lost and $Q(t|\alpha)\to Q(t)$.
We derive this stability in the linear approximation. 
Let us assume that $Q(t|\alpha)=\bar{Q}(t)+\beta(t|\alpha)$, where $\beta(t|\alpha)$ is a small transversal perturbation.
Then the equation for $\bar{Q}(t)$ is~\eqref{eq:eqQ}, and the equation for $\beta(t|\alpha)$ reads $\dot \beta=(i\omega_0-\gamma) \beta-2H^* \bar{Q}\beta$.
Let us introduce still another variable $Y$ according to $\beta=\bigl(1-|\bar{Q}|^2\bigr)Y$.
Then the dynamical equation for this new variable reads
$\dot Y\!=\!\Bigl(i\omega+H\bar{Q}^*-H^*\bar{Q}-\gamma \bigl(1+|\bar{Q}|^2\bigr)\bigl(1-|\bar{Q}|^2\bigr)^{-1}\Bigl)Y$.
Because $\text{Re}\bigl(H\bar{Q}^*-H^*\bar{Q}\bigr)\!=\!0$, for the absolute value of $Y$ we obtain
$\frac{d}{dt}|Y|=-\gamma\,\bigl(1+|\bar{Q}|^2\bigr)\bigl(1-|\bar{Q}|^2\bigr)^{-1}\,|Y|$.
Because according to~\eqref{eq:eqQ} $0\leq {|\bar{Q}|^2}^{\phantom{1}}\!\!\!\!<1$, variable $|Y|$ decays to zero exponentially, and thus $Y\to 0$.
This means that also $\beta\to 0$, which proves linear asymptotic stability.
One can see that a spread of natural frequency is essential; for $\gamma=0$, we have $|Y|=\mathrm{const}$, and the memory about initial conditions is not lost (this is another manifestation of the Watanabe-Strogatz integrability of a population of identical oscillators~\cite{Watanabe-Strogatz-93,Watanabe-Strogatz-94}).
 
After the convergence $Q(t|\alpha)\to Q(t)$, the dynamics of the population with a distribution of phase shifts reduces to a single dynamical equation for the auxiliary order parameter $Q(t)$, the original order parameters are related to it via circular moments of the distribution of $\alpha$:
\begin{gather}
\dot Q(t)=(i\omega_0-\gamma) Q(t)+H-H^* Q^2 (t)\;, \label{eq:eqQ2}\\
Z_m(t)=\eta_m Q^m(t)
\;.\label{eq:eqZ2}
\end{gather}
This completes a closed description for a population of oscillators with a first-harmonics coupling, Cauchy distribution of the natural frequencies, and an arbitrary distribution of the phase shifts.
We note that a similar to \eqref{eq:eqZ2} expression for $Z_1$ has been obtained in \cite{iatsenko2014glassy} for a steady state regime of the dynamics.

Above, we have argued for the GAS property for noisy oscillators with arbitrary coupling, and derived linear asymptotic stability for oscillators fulfilling the OA ansatz (i.e., with the first-harmonics coupling and the Cauchy distribution of natural frequencies).
In both cases, the dissipativeness leading to GAS is due to disorder, either in the form of noise or in the form of a continuous distribution of natural frequencies. The necessity of disorder is rather obvious because, in particular, for a first-harmonics coupling without disorder (i.e., for a population of identical oscillators), the Watanabe-Strogatz integrability holds, which prevents asymptotic stability.
However, the presence of disorder in the form of a continuous distribution of natural frequencies is generally not sufficient for GAS. On the one hand, a continuous distribution of natural frequencies leads to an effective (non-chaotic) mixing in a population due to the mechanism of Landau damping~\cite{Strogatz-00,fernandez2016landau,Dietert-16} (for explicit finite-dimensional dissipative reductions for distributions with nice analytic properties see~\cite{Klinshov_etal-21,campa2022study,Pyragas-Pyragas-22}).
On the other hand, this damping occurs only for differentially rotating oscillators and not for the locked ones.
The latter do not mix but form a coherent cluster.
One can expect GAS for the whole system if there can be just one cluster, i.e., if the coupling function $\Gamma(x)$ in \eqref{eq:dk} allows for one stable synchronous state for coupled units.
This is the case if this coupling function is just $\sin(x)$, like in the case where the OA ansatz is applicable.
Thus we expect that GAS will be valid for coupling functions possessing only one stable synchronous state but will be violated for multi-stable couplings~\cite{Daido-95,Komarov-Pikovsky-13a,Komarov-Pikovsky-14}.
Note that with an unbounded noise (e.g., with a Gaussian white noise assumed in the discussion of the Fokker-Planck equation above) the multi-stability disappears because now transitions between different stable clusters become possible. But for a bounded noise, multistability can survive. 

\begin{figure}
\centering
\includegraphics[width=\columnwidth]{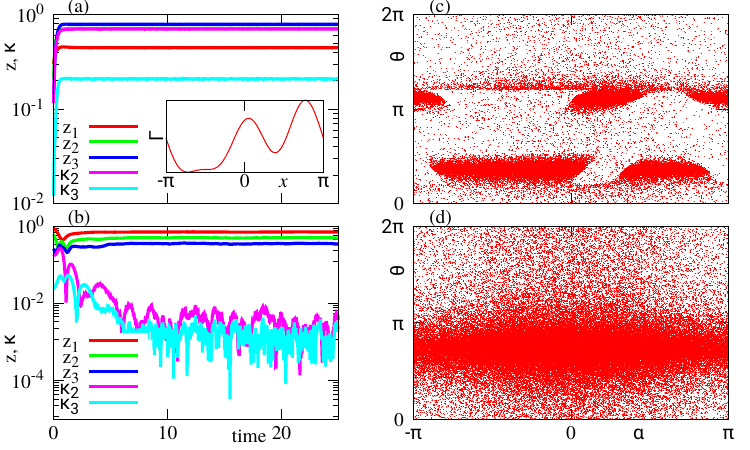}
\caption{Illustration of non-ergodicity for an ensemble of $N\!=\!10^{5}$ oscillators with a Cauchy distribution of natural frequencies and coupling function $\Gamma(x)$ having three harmonics (shown in the inset of panel (a)). 
Panels (a,b): evolutions of the moments $Z_m$ and the cumulants $\kappa_m$: in the case (a) cumulants attain large values indicating non-validity of the OA ansatz, while in case (b) they decay.
The corresponding distributions are illustrated by plotting states of all oscillators on the $\alpha,\theta$ plane with red dots.
Random variables $\theta$ and $\alpha$ are independent in panel (d) but show a complex pattern in panel (c).}
\label{fig:fig1}
\end{figure}

There is still an interesting possibility for GAS to be valid, at least for some initial states, for continuous distributions of natural frequencies even if the coupling $\Gamma(x)$ is so complex to beget multistability. 
This relies on the observation above that in the presence of GAS, the effective coupling function $\tilde\Gamma(x)$ appears, which is a convolution of the original coupling function and the distribution of the phase shifts.
Suppose that the distribution density of the phase shifts has the form \eqref{eq:g1h}, i.e., it has only one Fourier harmonics.
Then the effective coupling function $\tilde\Gamma(x)$ will also have only one Fourier harmonics, fulfilling conditions for the OA ansatz.
Thus, this system will possess the state described by Eqs.~\eqref{eq:eqQ2},\eqref{eq:eqZ2} above (provided the distribution of frequencies is a Cauchy one).
This regime is expected to be robust, so it has a finite basin of attraction even if the initial conditional distribution of the auxiliary phases $\theta=\varphi+\alpha$ does depend on $\alpha$.
However, this state may coexist with other states, not fulfilling GAS. 
We illustrate this by considering a population of oscillators with a Cauchy distribution of natural frequencies, one-harmonic distribution of the phase shifts~\eqref{eq:g1h}, and a rather complex original coupling function $\Gamma(x)$ (see inset in Fig.~\ref{fig:fig1}(a)).
We show two runs from different initial phase distributions; in one (panels (b,d)), the system converges to a state where the distribution of $\theta$ does not depend on $\alpha$, while in another run (panels (a,c)) this dependence does not disappear. In the former state, the theory developed above can be applied.
Because the effective coupling function $\tilde\Gamma(x)$ possesses only one harmonics, the final state lies on the OA manifold.
To demonstrate this, we together with circular moments $Z_{m}$, $m=1,2,3$, show the two circular cumulants $\kappa_{2}=Z_{2}-Z_{1}^{2}$ and $\kappa_{3}=\frac{1}{2}(Z_{3}-3Z_{2}Z_{1}+2Z_{1}^{3})$, which characterize deviations from the OA manifold~\cite{Tyulkina_etal-18}.
One can see that over time, the magnitudes of these cumulants decrease (and saturate at the level of the finite-size fluctuations). In contradistinction, for other initial conditions, a regime with a clear dependence on the distribution on $\alpha$ establishes (see panel (c)), in which the effective coupling function cannot be introduced, and the state is far from the OA manifold.

Summarizing, we have demonstrated that in many situations, the dynamics of an ensemble of globally coupled oscillators with random distributed phase shifts can be reduced to an effective ensemble without phase shifts, with a proper reduction of the effective coupling function (the convolution of the original one and the distribution of the phase shifts).
This reduction heavily relies on a possibility to average over the phase shifts. For systems with maximal disorder where the phase shifts constitute a random matrix, this property appears to be justified in the thermodynamic limit.
For a partial disorder, where the matrix of phase shifts has constant raws or columns, we rely on the GAS property, ensuring convergence to the same asymptotic state (potentially non-stationary) from arbitrary initial distributions, provided that the time evolution of driving forces on the oscillators is fixed.
This property of independence of the final state on the initial distribution can be recast as ergodicity.
The GAS property can be naturally expected for ensembles with unbounded independent noises (e.g., Gaussian or Cauchy noise) and \textit{any} coupling function. 

The situation is more subtle if there is no noise, but a disorder is due to a continuous distribution of natural frequencies. Here one can expect GAS only for simple enough coupling functions, possessing no multistability.
For example, for a single-harmonic coupling function and a Cauchy distribution of natural frequencies, we have explicitly demonstrated asymptotic stability by virtue of the OA approach.
Furthermore, we showed numerically that the GAS property might be violated for cases where the coupling function possesses multistability. 

In our setup, we assumed that the phase shifts are random parameters of oscillators, independent of the natural frequencies and other parameters that can potentially be distributed (e.g., noise strengths). Another idealization adopted in the present study is that the magnitude of the coupling is the same for all units.
The latter assumption might be violated if the propagating global field decays on the way to remote units.
Our theory is only valid if this effect is much smaller than the spread of the phase shifts. 
This happens if the wavelength of the signal-transmitting wave is small enough.

\acknowledgments
The authors thank K. Khanin, E. Knobloch, J. Winkler, and S. Zelik for valuable discussions. LAS was supported by the Russian Science Foundation (grant no. 22-12-00348).

%

\end{document}